# Removing Degeneracy of Microlensing Light Curves Through Narrow-Band Photometry of Giants

Abraham Loeb and Dimitar Sasselov[1]
Harvard-Smithsonian Center for Astrophysics
60 Garden St., Cambridge MA 02138

## ABSTRACT

The standard light curve of a microlensing event provides only two constraints on the six unknown parameters of the lens. We show that narrow–band photometry during a microlensing event of a giant star can in addition determine the angular radius of the Einstein ring and the proper motion of the lens. This possibility results from the fact that the extended atmospheres of giants emit the cores of resonant lines primarily from a narrow ring (limb brightening). The radius of the emission ring can be determined to a precision of $\pm 20\%$ by detailed spectroscopic observations of the source after the lensing event has ended. A considerable fraction of the clump giants in the bulge have a ring radius $\gtrsim 10^{12}$ cm, within the range of Einstein radii for sub-solar mass lenses. The extended thin ring also provides a sensitive probe of possible planetary companions as it sweeps across the lens plane. The ring signature can be detected photometrically, using a narrow–band filter centered on the CaII K line at 3933Å.

*Subject headings:* astrometry–dark matter–gravitational lensing

---

[1] Hubble fellow

## 1. INTRODUCTION

By now there are more than 50 microlensing events detected towards the Large Magellanic Cloud (LMC) and the Galactic bulge (Alcock et al. 1995; Auborg et al. 1993; Udalski et al. 1994a). The standard light curve of an event is characterized by two observables: its peak amplitude and duration (Paczyński 1986). However, there are six unknown parameters that define the physical properties of the lens, namely its mass, distance, its three velocity components, and the impact parameter of its trajectory relative to the line of sight to the source. The mass, distance, and transverse velocity of the lens combine to set the duration of the event, and together with the impact parameter they also define its peak amplitude. Breaking the degeneracy among these parameters is of fundamental importance to the interpretation of microlensing light curves. For observations towards the LMC, it is essential to discriminate between lenses that are located within the LMC (Sahu 1994) and lenses which are part of the Galactic halo or thick disk (Gould, Miralda-Escudé & Bahcall 1994); for bulge observations, it is important to separate lenses that are embedded in the Galactic disk from lenses inside a possible Galactic bar (Zhao, Spergel, & Rich 1995; Han & Gould 1995). Removal of the existing degeneracy between the lens parameters is a necessary step towards a better understanding of the nature of the lenses. A variety of methods to partially achieve this goal were suggested recently. The most ambitious proposal involves a parallax satellite mission to measure the lens proper motion (Refsdal 1966; Gould 1992, 1994b, 1995a). Other proposals consider alterations of the standard light curve due to different properties of the source, such as its finite size (Gould 1994a, 1995b; Nemiroff & Wickramasinghe 1994; Witt & Mao 1994), its polarized emission (Simmons, Newsam, & Willis 1994), spectral shifts due to its rotation and finite size (Maoz & Gould 1994), or shifts in its color due to a luminous lens (Kamionkowski 1995; Udalski et al. 1994c). In this *Letter*, we show that narrow–band photometry of lensed giant stars can be used to measure the proper motion and the angular radius of the Einstein ring of the lens.

Optical starlight is dominated by photospheric continuum emission. The continuum emission generally peaks at the projected center of the stellar disk and falls–off towards the stellar limb. In cool stars, some spectral lines (such as H$\alpha$ or Ca II) are still optically thick at the limb and therefore appear strongly in emission there. This resonant line scattering was predicted for the Sun in 1870 by C.A. Young and was later observed during solar eclipses (Zirin 1988). The limb brightening of various strong lines in the Sun extends only over a ring of width $\lesssim 0.01 R_\odot$. Cool stars with a much smaller surface gravity, such as giants and supergiants, have atmospheres which are more extended than the Sun by 1-2 orders of magnitude and therefore show a much stronger resonant line scattering. Limb brightening of supergiants has been observed directly only recently (Hannif et al. 1992).

For giant stars, the line emission ring has a radius of 0.05-1AU, comparable to the Einstein radius of sub-solar mass ($\lesssim 0.1 M_\odot$) lenses. A ring of this size provides just the baseline necessary to mimic observations of a microlensing parallax by an Earth–satellite system. For a specific star, the actual radius of the ring can be measured up to a precision of $\sim 20\%$ after the lensing event has ended, using traditional methods of high-resolution spectroscopy and infrared photometry (see McWilliam & Rich 1994 for application to Galactic bulge giants). Based on the measured radius, it is then possible to determine the projected Einstein radius and the proper motion of the lens relative to the line of sight. The extended thin emission ring also provides a sensitive probe of planetary companions as it sweeps across the lens plane.

The outline of this paper is as follows. In §2 we present the method used for calculating the narrow–band light curves of giants. We calculate disk intensity distributions for several atmospheric models and a 6–level Ca II atom model. In §3 we estimate the probability distribution of radii for clump giant stars in the Galactic bulge, based on the color–magnitude diagrams of the OGLE observations (Stanek et al. 1994). Numerical examples for narrow–band light curves of bulge giants are presented in §4. Finally, §5 summarizes the main conclusions of this work.

## 2. NARROW-BAND LIGHT CURVES OF GIANTS

The amplification of a point source by a point mass, $M$, depends only on their projected separation $d$ (Paczyński 1986),

$$A(d) = \frac{d^2 + 2}{d(d^2 + 4)^{1/2}} \quad (1)$$

where $d$ is expressed in units of the angular radius of the Einstein ring of the lens, $\theta_E = ([4GM/c^2][D_{\rm LS}/D_{\rm OL} D_{\rm OS}])^{1/2}$, and $D_{\rm OL}, D_{\rm LS}$, and $D_{\rm OS}$ are the distances between the observer, lens and source. The total flux received from an extended source is therefore obtained by integration over its infinitesimal elements,

$$F(t) = \int_0^{\tilde{R}_s} r dr B(r) \int_0^{2\pi} d\theta A(d), \quad (2)$$

where $B(r)$ is the surface brightness profile of the source in the projected polar coordinates $(r, \theta)$ around its center. For simplicity, we assume that the stellar emission is circularly symmetric, and denote the source radius by $R_s = \tilde{R}_s \times (\theta_E D_{\rm OS})$. All projected length scales are normalized by the Einstein ring radius. Using the law of cosines

$$d = |\vec{d}_0 - \vec{r}| = (d_0^2 - 2d_0 r \cos\theta + r^2)^{1/2} \quad , \quad d_0 = (b^2 + v^2 t^2)^{1/2}, \quad (3)$$

where $d_0$ is the projected separation between the lens and the source center, $b$ is the projected impact parameter of the source center, $v$ is the transverse velocity of the lens in units of $\theta_E D_{\rm OL}$, and $t = 0$ is the midpoint time of the lensing event. For a circularly symmetric source, the resulting light curve is time-symmetric with $F(t) = F(-t)$.

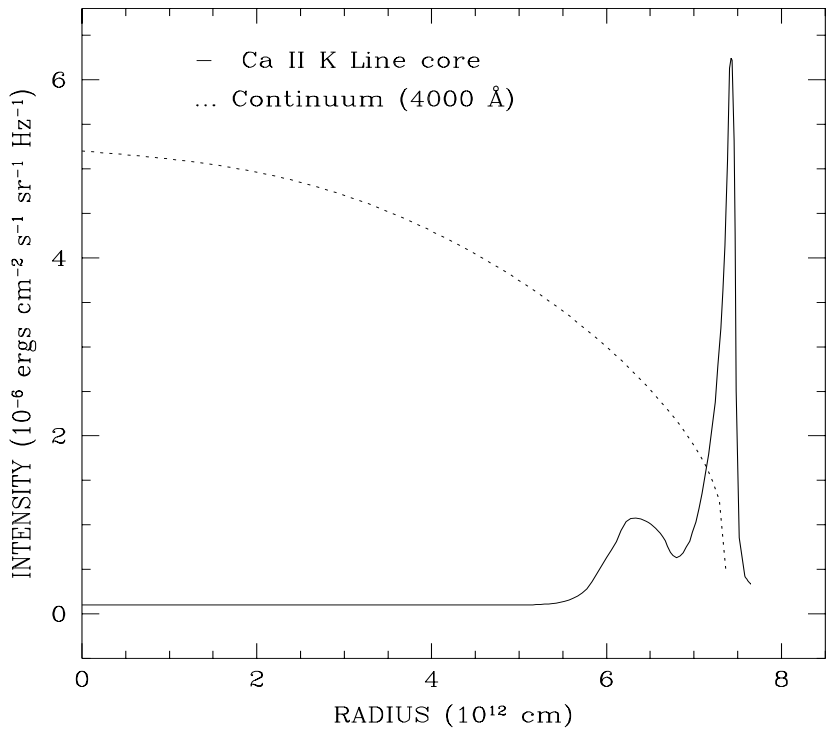

Fig. 1.— Stellar disk intensity distributions from a model atmosphere calculated in spherical geometry. The solid line shows the center-to-limb intensity variation in the core of the Ca II K $\lambda$ 3933 line for a 100 $R_\odot$ cool (4150 K) giant star. The dotted line shows the broad–band continuum intensity variation in the spectral region adjacent to the Ca II line.

The narrow–band light curve of a giant is obtained by substituting its center-to-limb intensity distribution, $B(r)$, in equation (2). In calculating $B(r)$ we ignore any surface features on the otherwise circularly-symmetric stellar disk, i.e. our radiative transfer models are one-dimensional. As initial conditions for constructing spherical atmospheres we use constant-flux plane-parallel blanketed photospheres from Kurucz (1993), and outer atmosphere models from Dupree, Hartmann, & Avrett (1984) and Harper (1992). The multi level non-LTE radiative transfer in these models is then solved in a spherical geometry using the SMULTI code (Harper 1994). We performed calculations with several atomic models which have strong observable resonance lines (such as H, Ca II, Mg II) and found that both the Ca II H & K lines show significant scattering above the limb, and are idealy suited for photometry. The model used for the Ca II atom has 6 levels; it treats both resonance lines and the infrared triplet in detail. A discussion of a similar calculation and a comparison with plane-parallel profiles can be found in Sasselov & Karovska (1994). An example for the most extended atmosphere we have considered is illustrated in Fig. 1, where we show the intensity distribution across a giant with a radius of $100 R_\odot$ and an effective temperature of 4150 K. The intensity distribution for the Ca II K line core reflects our optimal choice of a wavelength band. This filter passband was chosen based on our modeled line profiles and on practical considerations regarding existing observational setups–to be discussed in § 4.

## 3. GIANTS IN THE BULGE

The giants in the bulge have a relatively high metalicity (McWilliam & Rich 1994), implying a high opacity in their interiors. For a given stellar mass, a higher opacity favors convection and leads to a larger radius, $R_s$, and a cooler effective temperature (Iben & Faulkner 1968). The red giants in the bulge are similar to the members of the old open cluster M 67, whose color-magnitude relation is often used to represent the bulge (Ibata & Gilmore 1995). In order to determine the probability distribution of giant radii in the bulge, we use the color-magnitude diagrams provided by the OGLE collaboration for Baade's Window (Stanek et al. 1994; see similar analysis by Witt 1995).

The relation between absolute magnitude, color, and radius, can be established using direct measurements of giant radii in the solar neighborhood and through the calibration of the M 67 cluster. In particular, one needs to find an adequate temperature calibration for the Cousins V-I color used in the OGLE observations of the bulge. We determine the extinction based on the better known value for E(B–V) = 0.48 (e.g. McWilliam & Rich 1994), and the transformations of Taylor (1986) for K2 III stars. We find E(V–I) = 0.65. This result is in good agreement with the HST photometry of Baade's Window by Holtzman et al. (1993)–using the new abundance estimates by McWilliam & Rich (1994), as well as with the recent result from W UMa systems by Rucinski (1995). Next, we use as most reliable the direct temperature calibration of V–K photometry from lunar-occultation angular-diameters of field giants by Ridgway et al. (1980) and from Michelson interferometry by Di Benedetto & Rabbia (1987) and Mozurkewich et al. (1991). This calibration is transformed to match the bulge giants by using the giants in M 67 as surrogates with data from Taylor & Joner (1988) and Joner & Taylor (1988). As an independent check, we compared the recent spectroscopic temperatures of 11 bulge giants by McWilliam & Rich (1994) to the above calibration of the giant's effective temperature and found a very good match, except for three unusual cases.

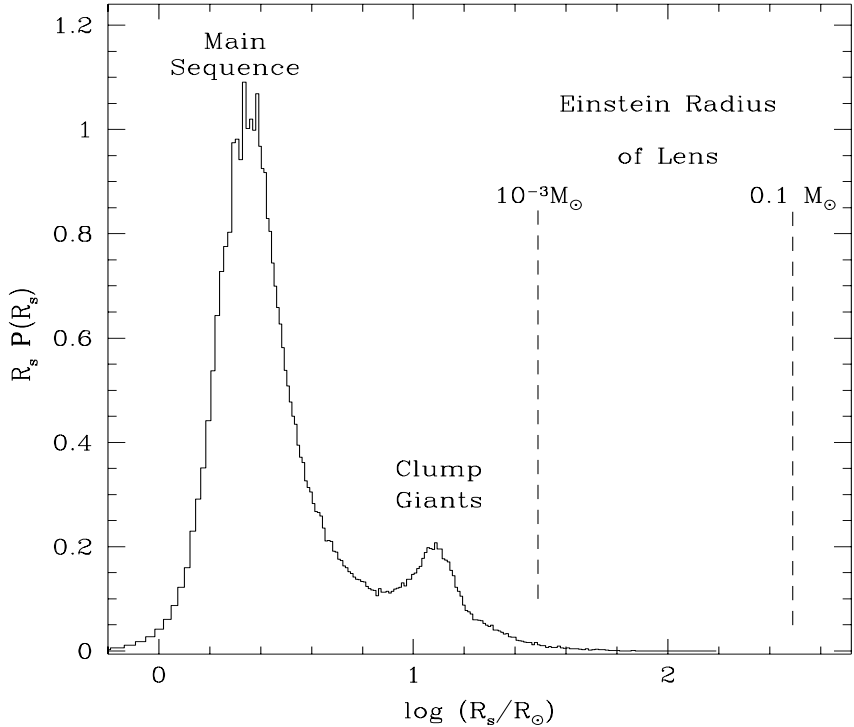

Fig. 2.— Probability distribution of stellar radii in Baade's Window, derived from the OGLE data set (Stanek et al. 1994). The clump giants are clearly separated from the Main Sequence. The Einstein radius for a lens mass of $10^{-3}$ $M_\odot$ ($32 R_\odot$) and a lens mass of 0.1 $M_\odot$ ($316 R_\odot$) are marked, assuming $D_{OS} = 8$ kpc and $D_{OL} = 6$ kpc. The Main Sequence is only qualitatively illustrated–our radius calibration applies to giants and supergiants.

Using the above temperature calibration and adopting a distance to the Galactic center of $D_{OS} = 8$ kpc, we have computed the radii of all stars for which the above relations are valid in the OGLE data set ($\sim 50\%$ of the entire stellar population). Figure 2 shows the probability distribution of stellar radii for $\sim 90\%$ of the OGLE stellar population (V–I $\geq 0.67$). Since we are interested primarily in cool giants and supergiants, we did not use a separate radius calibration for dwarfs; the Main Sequence is shown in Fig. 2 only to illustrate its relative fraction in the observed stellar population. We find that giants with radii $\gtrsim 20 R_\odot$ comprise 13% of the clump giants in Baade's Window. About 19% of all the stars have radii $\gtrsim 5 R_\odot$. The effect of a possible stellar bar on this distribution was found to be small and comparable to other uncertainties. However, since a significant fraction of the lenses may reside in such a bar (Zhao, Spergel, & Rich 1995), we adopt a distance to the lens of $D_{OL} = 6$ kpc and a transverse lens velocity of 100 km s$^{-1}$ in the numerical examples that follow.

## 4. RESULTS

The light curves of ring-like sources separate into two types: those with a single smooth peak and those with two sharp spikes. The first type is obtained when the lens is outside the ring trajectory, i.e. $b \gtrsim \tilde{R}_s$, and the second type is obtained when the lens crosses the ring, i.e. $b < \tilde{R}_s$.

For the extreme first-type curves with $\tilde{R}_s \ll b$, it is possible to solve equation (2) analytically. The amplification can then be Taylor expanded to second-order,

$$A(d) \approx A(d_0) + A'(d_0) \times (d - d_0) + \frac{1}{2} A''(d_0) \times (d - d_0)^2, \qquad (4)$$

where its derivatives are obtained from equation (1)

$$A'(d_0) = -\frac{8}{d_0^2 (d_0^2 + 4)^{3/2}} \; ; \; A''(d_0) = \frac{8(5d_0^2 + 8)}{d_0^3 (d_0^2 + 4)^{5/2}}, \qquad (5)$$

and where equation (3) gives,

$$(d - d_0) \approx -r \cos\theta + \frac{r^2}{2 d_0} \sin^2\theta \; ; \; (d - d_0)^2 \approx r^2 \cos^2\theta . \qquad (6)$$

The light curve of a point source, $F_0(t)$, is obtained by using $A(d_0)$ instead of $A(d)$ in equation (2). The fractional deviation from this light curve due to a source with a nonzero size is obtained by substituting equations (4)–(6) into equation (2),

$$\Delta(t) \equiv \frac{(F - F_0)}{F_0} = H(t) \langle r^2 \rangle , \qquad (7)$$

where $\langle r^2 \rangle \equiv [\int_0^{\tilde{R}_s} r dr \, B(r) \, r^2]/[\int_0^{\tilde{R}_s} r dr \, B(r)]$, and where the temporal profile of the excess signal is given by,

$$H(t) \equiv \frac{8(d_0^2 + 1)}{d_0^2 (d_0^2 + 2)(d_0^2 + 4)^2} . \qquad (8)$$

For emission from a narrow ring, $\Delta(t) = G(t) \tilde{R}_s^2$. Thus, common events with $b \sim 0.5$ yield maximum values of $\Delta \sim \tilde{R}_s^2$, so that a 3% accuracy in the photometry of the light curve which includes ten points of observations can reveal the signature of a source radius $\tilde{R}_s \gtrsim 0.1$. For our lens-source parameters, this threshold corresponds to a source radius $\gtrsim 2 \times 10^{12}$ cm $\times (M/0.1 M_\odot)^{1/2}$, which is in the expected range of radii for clump giants in the bulge (cf. Fig. 2). Based on a comparison to the exact solution, we find that equations (7) and (8) provide a good description to the ring light curve as long as $\tilde{R}_s < b$.

However, the character of the light curve changes for $\tilde{R}_s > b$, when the ring is able to cross the lens. In this case, the event contains two pronounced peaks which are centered around the two crossing times of the ring through the lens. These crossing times are defined by the condition $d = |\vec{d}_0 - \vec{R}_s| = 0$. The structure of the peaks depends on $\tilde{R}_s$; a larger $\tilde{R}_s$ makes the peaks narrower because the fraction of the ring area that is being affected by the lens gets smaller. By fitting the peak structure and separation to a stellar atmosphere model, it is possible to find both the impact parameter $b$ and the source size $R_s$ in units of the Einstein radius.

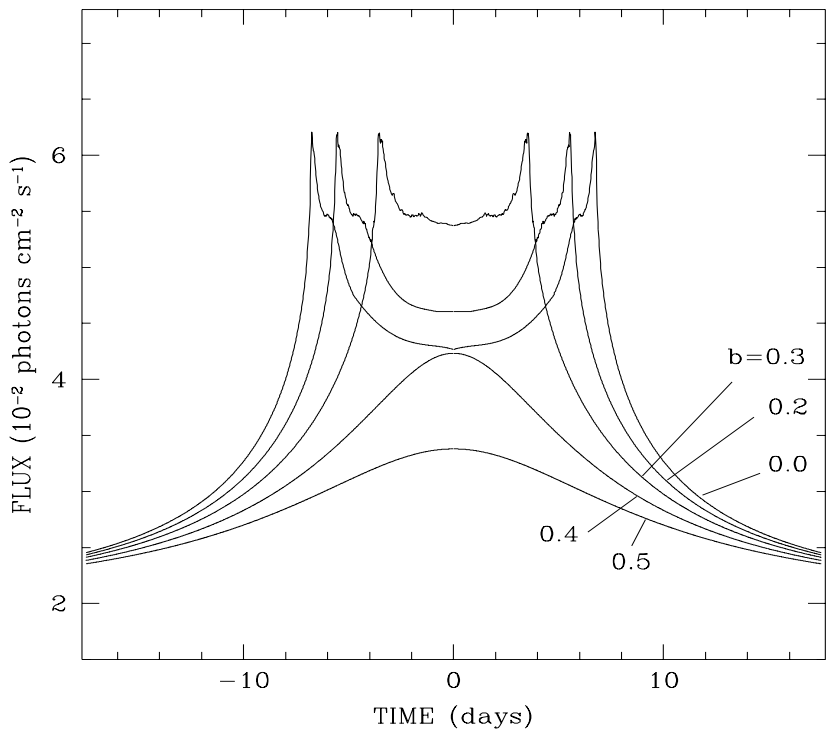

Fig. 3.— Flux amplification profiles in a narrow passband centered on the Ca II K line for lensing of the giant shown in Fig. 1 by a $0.1 M_\odot$ mass lens. The light curves are shown for different values of the dimensionless impact parameter, $b$. The radius of the Ca II ring of the giant is $\tilde{R}_s = 0.35$ in Einstein ring units. We plot the net photon flux on Earth for the spectrophotometry filter described in § 4.

Figure 3 presents the distribution of narrow–band light curves as a function of the impact parameter $b$ for a lens mass $M = 0.1 M_\odot$ positioned at $D_{\rm OL} = 6$ kpc and moving with a transverse velocity of 100 km s$^{-1}$ relative to the line of sight to a clump giant source with $R_s = 100 R_\odot$ (cf. Fig. 1) at $D_{\rm OS} = 8$ kpc. As this figure shows, it is much easier to identify the signal of the ring at small impact parameters. Figure 4 compares the narrow–band (solid line) and the broad–band (dashed) light curves of this source for a lens mass of $10^{-3} M_\odot$ and $b = 0$. The main difference between the curves is the existence of two peaks in the narrow–band curve. The significance of these peaks is emphasized by the similarity between the broad–band curve (dashed) and a standard microlensing curve for a point source with some different impact parameter and velocity (dotted). Although the small difference between these light curves is detectable, the extraction of the value of $\tilde{R}_s$ from this difference could be complicated by observational errors or blending with unlensed starlight in crowded fields (Di Stefano & Esin 1995). The finite source signature can be measured more reliably and unambiguously from the two peaks in its narrow–band light curve. For $D_{\rm OL} = 6$ kpc, the fraction of giant lensing events with $b < \tilde{R}_s$ is of order $\langle \tilde{R}_s \rangle \approx 4\% \times (M/0.1 M_\odot)^{-1/2}$ (cf. Fig. 2).

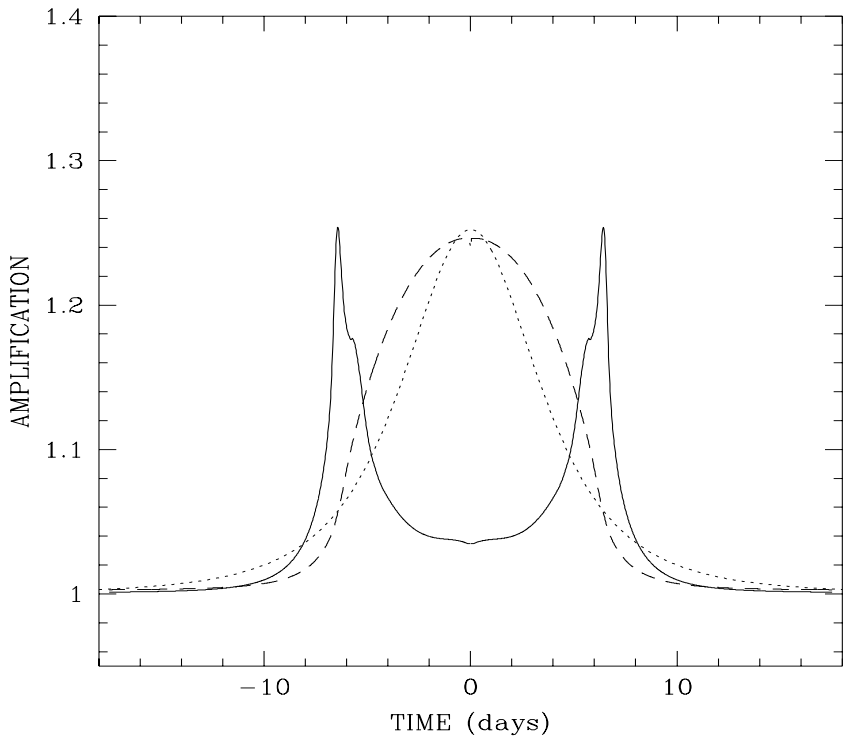

Fig. 4.— Comparison of light curves for a lens mass of $10^{-3} M_\odot$. The narrow–band light curve (solid line) is compared to the 4000Å broad–band light curve (dashed), for the giant shown in Fig. 1 with $\tilde{R}_s = 3.5$ and $b = 0$. The dotted line is a standard microlensing curve for a point source, chosen to fit closely the broad–band curve; it corresponds to $b = 1.15$ and a transverse lens velocity of 50 km s$^{-1}$.

The several decades of experience in flux measurements of Ca II emission from dwarf and giant stars should ease the technical aspects of the narrow–band photometry required here. In practice, one could use a modern version of the four-channel photon-counting spectrophotometer (*HKP2*) of the Mt. Wilson HK project (Duncan et al. 1991) with 20Å–wide reference bands and 2Å–wide H and K bands. For example, we find that the net flux in these H and K passbands per unit area at the stellar surface is $4 \times 10^5$ erg cm$^{-2}$ s$^{-1}$, for a giant with a radius of 20 $R_\odot$ and an effective temperature of 4400 K (see Rutten (1984) for absolute flux calibration in these passbands). With the adopted distance to the bulge, this translates to a detection rate of about 1 photon/sec on a 2-meter telescope. Therefore, high quality ($\pm 2\%$) narrow–band photometry can be achieved using the *HKP2* instrument with an integration time of $\sim 45(D_{\rm t}/2 \text{ m})^{-2}$ minutes on a telescope with a diameter $D_{\rm t}$. This result applies to the signal without amplification; the peaks of the light curve shown in Fig. 4 are easier to observe. In addition, giants with $R_s > 20R_\odot$ require much shorter integration times as can be seen from Fig. 3. Occasionally, a microlensing event may involve a chromospherically active giant which happens to have an active phase at the time of the event. Such activity could distort the circular symmetry of the emission ring or add bright spots on the stellar disk. However, subsequent photometry would be able to determine whether the lensed star was likely to have been active during the event.

## 5. CONCLUSIONS

A considerable fraction of the giants in the bulge have a radius $R_s \gtrsim 10^{12}$ cm (cf. Fig. 2). As illustrated in Fig. 1, the narrow–band (resonance line) images of these giants are thin rings. When a ring-like image crosses a lens, its time–symmetric light curve shows two sharp peaks. These narrow peaks provide an unambiguous signature of the finite source size. The broad–band light curve contains a smooth, weaker signature of the source size (cf. Fig. 4), that could be inaccurately interpreted due to observational errors or blending with unlensed starlight in crowded fields. The radius of lensed giant can be measured spectroscopically to within $\pm 20\%$ after the lensing event has ended. The measured radius can then be combined with the narrow–band light curve to infer the proper motion and the angular radius of the Einstein ring of the lens. These determinations can be used to discriminate between the different distance scales possible for the lens population.

The existence of an extended thin emission ring enhances the probability of detecting low-mass companions of the lens, such as planets (for discussions about planetary lensing of a point source, see Mao & Paczyński 1991, Gould & Loeb 1993, and Bolatto & Falco 1994). First, the finite source size sweeps through a wide strip in the lens–companion plane. Thus, the probability of intersecting the lensing region of influence of a companion planet is increased relative to the case of a point source by the ratio between the source size $R_s$ and the Einstein radius of the planet. For bulge observations, this ratio is $\sim (R_s/2 \times 10^{11} M_{\rm pl}^{1/2}$ cm), where $M_{\rm pl}$ is the planet mass in earth mass units. Second, the planetary amplification is not averaged–out by the entire stellar disk as in the case of broad–band observations, but only by the width of the narrow emission ring–which is an order of magnitude smaller (cf. Fig. 4). Even in the extreme case of Fig. 1 where $R_s = 100 R_\odot$, the width of the emission ring is still smaller than the Einstein radius of a Jupiter-mass planet $\sim 2 \times 10^{12}$ cm. The ring provides two chances for the detection of a planetary–crossing signature. Based on the timing of these crossings relative to the primary lens crossings, it is possible to infer both the angular separation and the orientation of the companion relative to the primary lens. The proper motion of the two objects can be compared for any difference associated with their relative orbital motion[2]. Each planetary crossing signature may be multiply peaked and is expected to last $\lesssim$ day (cf. Gould & Loeb 1992). Once a lensing event of a giant star has triggered one of the currently operating early-warning systems (e.g. Udalski et al. 1994b), it would be useful to monitor its light curve on an hourly basis with a narrow–band filter. This may prove to be an efficient method in searching for planetary systems around stellar lenses.

We thank R. Di Stefano and S. Mao for useful discussions.

---

[2]This approach can also be applied to binary systems with lenses of comparable masses, although the light curves of these systems cannot always be decomposed into a brief disturbance on top of an otherwise standard curve (Mao & Di Stefano 1995; Udalski et al. 1994c).